\documentstyle[12pt,twoside]{article}
\begin{document}


\title{\small \rm \begin{flushright} \small{hep-ph/9511314}\\
\small{RAL-95-064}\\
\small{WIS-95/57-Nov.-Ph} \end{flushright} \vspace{2cm}
\LARGE \bf A Strong QCD Enhancement of CP Violation in Charmed Meson
Decays?\vspace{0.8cm} }
\author{Frank E. Close\thanks{E-mail : fec@v2.rl.ac.uk}\thanks{Work
supported in part by the European Community Human Capital Mobility program
``Eurodafne", Contract CHRX-CT92-0026.} \\
{\small \em Particle Theory, Rutherford-Appleton Laboratory, Chilton,
Didcot OX11 0QX, UK} \\ \\
Harry J. Lipkin \thanks{E-mail : FTLIPKIN@ WEIZMANN.WEIZMANN.AC.IL} \\
{\small \em
Department of Particle Physics,Weizmann Institute of Science,
Rehovot 76100, Israel}
\\
\centerline{and}
\\
{\small \em School of Physics and Astronomy,
Tel Aviv University,Tel Aviv, Israel}
    \\  \\}
\date{November 1995 \vspace{1.5cm}}

\begin{center}
\maketitle

\begin{abstract}
 The mass difference of $c$ and $u$ quarks is similar to the energy needed to
excite gluonic degrees of freedom in strong QCD. This underpins the degeneracy
of $D$ and a recently observed candidate hybrid excitation of the $\pi$ at$\sim
1.8$GeV
and may generate a strong analogue of Penguin mixing together with a measurable
CP-violation in $D$ and $D_s$ decays.
\end{abstract}
\end{center}

\newpage

The asymmetric decays of K-mesons and, possibly, our existence
(via the baryon antibaryon asymmetry of the universe)
are to date the only direct evidence for CP-violation.
The confirmation of three generations
of quarks and leptons encourages the suspicion that CP violation may be
intimately related to the presence of a phase in the CKM matrix
and be large in the third generation $B$-system.
In this standard model the charm sector is not expected to show significant
CP violation effects, the asymmetry for the $D^0 \rightarrow K^+K^-$ mode
being at most a few $10^{-3}$\cite{bigi}.

We point out here that the recent confirmation of a nonstrange $0^{-+}$ meson
resonance  with the quantum numbers of a pion\cite{ves,bnl95,bellini,pdg94}
 and, at 1.8GeV mass, in the
vicinity of the $D$-mass, raises the possibility that $D$ decay
branching ratios may be influenced by the presence of this resonance.
Furthermore, hints that this resonance is a hybrid\cite{close94,cp94,paton85}
(where the gluon degrees of freedom are excited in the presence of ground
state $q\bar{q}$) degenerate with the $D$ may enhance the gluonic
penguin diagrams
in $D$-decays and lead to observable CP violating effects in channels such
as $\pi\pi\pi$, $\pi K \bar{K}$, $\pi \eta\eta$ in $S$-wave which $D$ and
$\pi(1.8)$ share in common. Such an eventuality
would have far reaching implications: it would provide a new window into CP
violation, and also help to shed light on the dynamics of strong QCD which
remains one of the least understood parts of the standard model.

The large
number of open channels available for Cabibbo-suppressed $D$-decays into
nonstrange final states enable conclusive experimental investigations. The
first step is to examine the relative branching ratios of the $ D$ and this
$\pi(1.8)$
resonance to see if there are any common unexpected systematics. There are not
yet enough data for any reliable conclusions, but it is interesting that
in both cases the $3 \pi $ decay mode shows no signal for
$\rho \pi$  and may be
dominated by final states in which two of the pions are in
$S$-wave. The dominant decay modes observed for the $\pi(1.8)$ resonance are in
$\pi\pi\pi$ and $(\pi K\bar{K})_s$ with significant contribution
from $\pi f_0(1300)$ and $\pi f_0(980)$ (where $f_0(1300)$ is the broad
$\pi\pi$ $S$-wave structure discussed in ref.\cite{pdg94}).
So far $D^+ \rightarrow \pi^+ \pi^+ \pi^-$ has been measured, but its small
branching ratio  $\approx 3 \times 10^{-3}$ and limited
statistics means
that there has not yet been a conclusive Dalitz plot analysis;
(we are informed that this analysis is now in progress and hopefully data will
be available soon\cite{appel}).
The $\pi \eta \eta$ and $\pi \eta \eta'$ decay modes
 for the $\pi(1.8)$ resonance
have been seen\cite{ves91,rya}\footnote{This state is listed
in the Particle Data Tables as
$X(1830)$ with $J^{PC} = 2^{-+}$ preferred. However, it is
now established as $0^{-+}$\cite{ves,rya} and identical to the
$\pi(1770)$ of refs\cite{bellini,pdg94}},
the $\eta \eta \pi^-$ decays occurring at a level of $50 \pm10\%$
 of $\pi^+ \pi^- \pi^-$\cite{rya}.
This mode has not yet been looked for
in $D$ decay. If it is there also at a level of $\sim 50\%$
 of $3\pi$, it could be
sought at CLEO and provide further information on the relation between
D decays and the $\pi(1.8)$ meson resonance.

{}From the hadron spectroscopy point of view, the possible existence of
socalled ``exotic" hadrons like glueballs and hybrids is still an exciting
open question \cite{cafe,lattice,bes}.
 Radiative $J/\psi$ decays have been suggested as a good channel
for glueball searches  because they produce hadrons via a gluonic intermediate
state. Similarly heavy quark meson decays via penguin diagrams could provide a
good channel for hybrid (quark-antiquark-gluon) mesons because they produce
hadrons via a quark-antiquark-gluon intermediate state. The penguin
contributions
have generally been considered to be negligibly small in charm decays. However,
if the penguin contribution is enhanced by the presence of a hybrid
resonance near the $D$ mass, there may be an
appreciable penguin-tree interference
contribution which might also have a CP-violating relative phase.
Using the relative strengths of penguin and tree contributions predicted by the
CKM matrix elements, a very large enhancement of the penguin would be needed to
give observable CP-violating effects. However, the relative strength
predictions
have not yet been confirmed by experiment, and there may be additional
enhancements resulting from physics beyond the standard model. Thus at this
stage we keep all options open and concentrate on ways of using available data
to search for evidence for influence of the resonance on D decays.

We now examine in detail the contributions of
the Cabibbo-suppressed tree and penguin diagrams to $ 3 \pi $ and
$ K \bar K \pi   $ final states. For the $ 3 \pi $ final state the tree
diagrams
give
\begin{equation}
D(c \bar q_s) \rightarrow d + W^+ + \bar q_s \rightarrow
(d u \bar d) \bar q_s \rightarrow 3 \pi
\end{equation}
where the relevant weak vertex is
$ c  \rightarrow d + W^+ \rightarrow
(d u \bar d) $
and $\bar q_s$ denotes a spectator $\bar u$ or $\bar d$.
The penguin diagrams give

\begin{equation}
D(c \bar q_s) \rightarrow
b + W^+ + \bar q_s  \rightarrow u \bar  q_s G
\rightarrow 3 \pi
\end{equation}
where the relevant weak vertex is
$ c  \rightarrow b + W^+ \rightarrow u \bar q_s G$ and $G$ denotes a gluon.
The relevant weak vertices for the Cabibbo suppressed tree transition and the
penguin transition are seen to involve two different combinations of CKM matrix
elements. Thus the tree-penguin interference term can exhibit CP violation in
the standard model.

For the $ K \bar K \pi   $ final state the tree diagrams give
\begin{equation}
D(c \bar q_s) \rightarrow s + W^+ + \bar
q_s  \rightarrow
(s u \bar s) \bar q_s \rightarrow K \bar K \pi
\end{equation}
where the relevant weak vertex is
$c  \rightarrow s + W^+ \rightarrow
(s u \bar s) $.
 The penguin diagrams give
\begin{equation}
D(c \bar q_s) \rightarrow
b + W^+ + \bar q_s \rightarrow u \bar q_s G
\rightarrow K \bar K \pi
\end{equation}
where the relevant weak vertex is
$ c  \rightarrow b + W^+ \rightarrow u \bar q_s G.
$
Here again two different combinations of CKM matrix elements arise and the
interference term can exhibit CP violation.
The penguin contributions to both $ 3 \pi $ and $ K \bar K \pi $ final states
are seen to arise from the same relevant weak vertices, while the tree
contributions to  $ 3 \pi $ and $ K \bar K \pi $ can arise from different
relevant weak vertices. Thus the possibly CP-violating interference can be
different in the two cases and provide two different independent possibilities
for finding a CP-violating charge asymmetry.

The presence of a $0^{-+}$ hadron degenerate with the $D$ mesons requires
a reassessment of the CP studies for the $D$ in any event.
 In the quark model there are three possible sources
for a $0^{-+}$ state at such a mass:

(i) A $3S$ radial excitation of the $\pi$, predicted at $1.88$ GeV\cite{isgod}

(ii) A ``hybrid" state where the gluonic fields or ``flux tubes" are excited
in the presence of coloured quark sources, predicted at $1.8-1.9$
GeV\cite{paton85,bcs}

(iii) A $qq\bar{q}\bar{q}$ state\cite{suz,ter}

The effects of the presence of this resonance can be described in two
equivalent ways, since only effects of first order in the weak interaction
need be considered. One can consider the initial state as an unperturbed
charmed meson, calculate its decay as a first order perturbation in the weak
interaction using the weak interaction diagrams of the standard model, and
introduce the resonance as a strong final state rescattering. One can instead
examine the mixing of the $\pi$ resonance into the initial state via the weak
interaction. Using the latter picture we note that
although mixing with a nearby $\pi$ resonance
is expected whatever the dynamical structure of the latter may be, the effect
on
the penguin amplitude may be more dramatic for a hybrid.
Mixing between $D(c\bar{d};1S)$ and $\pi(u\bar{d};3S)$ via the penguin is
likely to be suppressed by the orthogonality of $1S$ and $3S$ wavefunctions to
the extent that the gluon transfers little momentum; by contrast the essential
transition in the penguin diagram is $c \rightarrow u + G$ and hence excitation
of gluonic modes, leading to hybrid final states ($\pi_H)$
may be expected. Indeed, given that gluonic excitations are believed to
require about $1 GeV$\cite{lattice,bcs} and that there is also a
$\sim O(1GeV)$ energy gap between $c$ and $u,d$ flavours, resonance with
excited gluonic modes may be significant and the degeneracy between $D$ and
 $\pi_H$
natural.

In ref.\cite{cp94} the relative sizes of the quasi two-body
branching ratios for a hybrid $\pi_H$ at $\sim 2$GeV were
predicted to be (after summing over all charge combinations)

\begin{equation}
\pi f_0(1300) \sim 16; \pi f_2 \sim 2; KK^*_0 \sim 18; \pi \rho
\sim 2; KK^* \sim 1
\end{equation}
which are consistent with the data on $\pi(1.8)$\cite{ves} where the
$KK_0^*$ channel is threshold forbidden and manifested as $(KK\pi)_S$.
A prominent $KK_0^*$ is observed and
the experimentally known
strong affinity of $K\bar{K} \rightarrow f_0(980)$ also probably is
responsible for the coupling to $\pi f_0(980)$ with a strength comparable to
the channel $\pi f_0(1300)$. The latter will also feed $\pi \eta \eta$.
Thus the overall expectations for hybrid $0^{-+}$ are in line with
the data of ref.\cite{ves}.

The branching ratios anticipated for a $3S$ state at this mass
are not like these.
The presence of nodes in the $3S$ wavefunction enable one to suppress the
$\pi\rho$ and $KK^*$ channels, thereby ``imitating" the hybrid signature, but
the presence of a dominant $\pi f_0(1300)$ and $\pi f_0(980)$ appear to be
unique to the hybrid configuration.

It is interesting that there
appear to be anomalies in the $D$-decays which may be direct manifestations of
the $\pi(1.8)$.

The Cabibbo favoured decays into three particle final states such as $D
\rightarrow K \pi\pi$ are dominated by quasi two body vector-pseudoscalar
production
\cite{alder}. If spectator diagrams dominated the $D \rightarrow \pi\pi\pi$
and $D \rightarrow K\bar{K}\pi$ channels, one would expect a similar phenomenon
 here;
however the data are very different\cite{pdg94} favouring ``non-resonant" modes

\begin{equation}
D^{\pm} \rightarrow (\pi\pi\pi)_{non-res} = (2.5 \pm 0.7) \times 10^{-3};
D^{\pm} \rightarrow (\rho\pi) < 1.4 \times 10^{-3}
\end{equation}
The situation is more complicated in the $K\bar{K}\pi$
channels
\begin{equation}
D^+ \rightarrow (K^+K^-\pi^+)_{non-res} = (4.6 \pm 0.9) \times 10^{-3};
\end{equation}
\begin{equation}
D^{+} \rightarrow (\bar{K}^{*o}K^+) \rightarrow (K^+K^-\pi^+) =
(3.4 \pm 0.7) \times 10^{-3}
\end{equation}
\begin{equation}
D^{+} \rightarrow (\phi \pi^+) \rightarrow (K^+K^-\pi^+) =
(3.3 \pm 0.4) \times 10^{-3}
\end{equation}
If ``non-resonant" means ``S-wave" then these modes
may be the $\pi f_0$ and $K K^*_0$
channels of the $\pi_H(1.8)$ driving the Dalitz plot. To test this we urge a
search for other modes of the $\pi(1.8)$ among the $D$ decays driven
by $\pi f_0$ such as $\pi \eta\eta$ and $\pi (\pi\pi)_s(\pi\pi)_s$ in the
$5\pi$ final state together with $\pi f_0(980)$.

There are similar anomalies in the $D^0$ decays. The $\pi^+ \pi^- \pi^0$
channel is large: ref.\cite{bal} reports $ 1.6 \pm 1.1 \%$ while
 ref.\cite{acc} in $\pi A \rightarrow (3\pi) A$ detected $D$ by its decay
vertices and find
$3.9\%$ which they acknowledge to be far in excess of that expected for a mode
that is conventionally expected to be Cabibbo disfavoured. The strong direct
excitation of $\pi(1.8)$ is unlikely to be  a major
contaminant here though it merits further study as does  the possible
contribution of missing $\pi^0$. The ratio of
$\rho \pi  $/$(\pi\pi\pi)_s$ needs quantifying.

In view of these qualitative hints we urge that the possibility of measurable
CP-violation in these channels be considered. Thus decays of $D^+$ and $D^-$
separately should be studied e.g. to test if the branching ratios $D^{\pm}
\rightarrow \pi^{\pm} \pi^{\pm} \pi^{\mp}$ or of $D^{\pm}
\rightarrow \pi^{\pm} K_S K_S $ are identical. Similarly
$D^0 \rightarrow K^0 K^- \pi^+$ and $\bar{K}^0 K^+ \pi^-$ non-resonant channels
are superficially different\cite{pdg94} though with large errors.

The ratio of $5\pi$ relative to $3\pi$ rates is also interesting. The $D
\rightarrow 5\pi$ come from ref.\cite{acc} which is technically only an
 upper limit due to the possible contamination from extra $\pi^0$. Thus
it is necessary to quantify the ratio $\frac{D \rightarrow 5\pi}{D
\rightarrow 3\pi}$ and compare with that for the $\pi(1.8)$. Preliminary
data from VES suggest that this ratio is significantly less than
unity\cite{sasha}.

The $K\bar{K}\pi$ channels are of particular interest for both charged and
neutral $D$ because the $\phi \pi$
decay is forbidden by the OZI rule for an ordinary $q \bar q$ meson. It is
allowed for $D$ decays via the tree diagram. Since both $\phi \pi$ and
$\rho\pi$
are vector-pseudoscalar states, presumably related by SU(3) symmetry, the
difference between the $D$ decays into these two final states may offer
interesting clues.
Additional information can be obtained from the $K_S K_S \pi$ decay mode, since
the p-wave decay modes including $\phi \pi$ are forbidden and only even partial
waves are allowed. Here the $K\bar{K} \rightarrow f_0(980)$ may show up as an
enhancement near threshold, without being masked by $\phi \pi$ which may well
dominate the charged $K\bar{K}\pi$ modes.

Note that for the quasi-two-body neutral decays $D^o \rightarrow K^* \bar K$
the tree diagram produces only a $u \bar u$ pair and not a $d \bar d$.
This diagram can give only the charged mode
$D^o \rightarrow K^{*+} K^{-}$; the neutral mode
$D^o \rightarrow K^{*o} \bar K^{o}$ is forbidden. Both modes are allowed for
the penguin. However, the $\pi(1.8) \rightarrow KK^*$ is
suppressed\cite{ves,cp94} and so we expect $K^{*o}K^{o}$ to remain small.

Our arguments may also apply to the $D_s$ decays. It has been a long standing
puzzle that $D_s \rightarrow \pi^+ \rho^0$ is suppressed relative to
non-resonant $(3\pi)^+$. Ref.\cite{suz} has suggested that this is due to
$qq\bar{q}\bar{q}$ and Terazaki\cite{ter} has also considered the influence of
$0^{-+}$ mesons in this region. The annihilation diagram could occur in
the sequence $c\bar{s} \rightarrow c\bar{s}gg \rightarrow u\bar{d}gg$ and mix
the $D_s$ into $\pi_H$; in such a case we would again expect further channels
in common with $\pi(1.8)$. By analogy with the foregoing discussion, and to
the extent that we anticipate the mass difference $K_H(0^{-+}) - \pi_H(0^{-+})
\sim D_s - D_d$ there will be $D_s$ mixing with $K_H$ via penguins
that will distort $D_s$ decays. The analysis of hybrid decays in
ref.\cite{cp94}
when extended to $K_H$ leads one to expect suppression of $\pi K^*$ and
$\rho K$ relative to $K f_0$ and $\pi K^*_0$ which both feed
the $(K \pi \pi)_s$ ``non-resonant". If $K f_0(1500)$ is produced (analogous
to $\pi f_0(1500)$ for $\pi_H$) then $K + 4\pi$ and $K \eta\eta$ may be
present.

Verification of a consistent pattern in the Cabibbo disfavored modes of $D$
and $D_s$ that parallels those of $\pi_H$ and its anticipated partner $K_H$
would both support the idea that gluonic fields are excited in these $\pi$ and
 $ K$ resonances and open the way to CP-violation
in these resonance enhanced modes. We urge that searches be made
for these channels together with partial wave analyses of the 3 and
5-body channels and that quantitative comparison be made between
$D^0$ and $ \bar{D^0}$ and between $D^+$ and $D^-$
as new windows into CP-violation.

Note that no such effects are expected in the $B$ system as the necessary
$0^{-+}$ hybrids are far displaced in mass. Consequently $B$ decays should
behave canonically. Verification of the expected $B$ decay phenomenology
together with anomalous $D$ (and $D_s$) decays in channels common to
$\pi_H$ (and $K_H$) would be a rather direct
proof of the presence of strong gluonic excitation at around 1.8 - 2GeV mass.

\vskip 0.2in

We are grateful to J.Appel, C.Damerell and S.Stone for discussions

\end{document}